\documentclass[12pt,thmsa]{article}
\usepackage{amsfonts}

\input{tcilatex}

\begin{document}

\begin{center}
{\LARGE Curvature and torsion of implicit hypersurfaces }

{\LARGE and the origin of charge}

{\huge \bigskip }

\textbf{R. M. Kiehn}

69 Chemin St. Donat, 84380 Mazan, France

http://www.cartan.pair.com

rkiehn2352@aol.com

\vspace{1pt}

\vspace{1pt}
\end{center}

\begin{quote}
\vspace{1pt}\textbf{Abstract:} \ A formal correspondence is established
between the curvature theory of generalized implicit hypersurfaces, the
classical theory of electromagnetism as expressed in terms of exterior
differential systems, and thermodynamics. \ Starting with a generalized
implicit surface whose normal field is represented by an exterior
differential 1-form, it is possible to deduce the curvature invariants of
the implicit surface and to construct a globally closed vector density in
terms of the Jacobian properties of the normal field. \ When the closed
vector density is assigned the role of an intrinsic charge current density,
and the components of the normal field are assigned the roles of the
electromagnetic potentials, the theory is formally equivalent to an exterior
differential system that generates the PDE's of both the Maxwell Faraday
equations and the Maxwell Ampere equations. \ The interaction energy density
between the potentials and the induced closed charge current density is
exactly the similarity curvature invariant of highest degree (N-1) for the
implicit surface. \ Although developed without direct contact with M-brane
theory, these ideas of generalized implicit surfaces should have application
to the study of p-branes that can have multiple components and envelopes. \
The theory suggests that gravitational collapse of mass energy density
should include terms that involve the interaction between charge-current
densities and electromagnetic potentials. \ 

\vspace{1pt}
\end{quote}

\section{Introduction}

\subsection{An overview}

\vspace{1pt}\qquad The origin of charge has long been a mystery to physical
theory, perhaps even more illusive than the concept of inertial mass. \ A\
major objective of this article is to examine the conjecture that the
charge-current density of electromagnetism may have its origins in the
differential geometry and topology of curvature and torsion, in a sense
similar to the idea that mass density and gravity\ have their origins in the
concept of curvature. \ The curvatures of interest are not those generated
by a symmetric metric, but instead are those similarity invariants
associated with a generalized implicit hypersurface. \ The generalized
implicit hypersurfaces considered may not admit a global foliation as their
normal fields need not satisfy the Frobenius integrability conditions. \
Hence such generalized hypersurfaces can support topological torsion as well
as curvature.

An arbitrary 1-form of Action, $A_{0},$whose coefficient functions may be
considered as a set of electromagnetic potentials$,$ when suitably scaled,
can also play the role of the normal field to a generalized implicit
hypersurface. \ The closure of the exterior differential system, $%
F_{0}-dA_{0}=0$, always generates a system of PDE's which contain the
Maxwell-Faraday equations. \ When the 1-form of Action is rescaled by use of
a Holder norm, $\lambda ,$ such that resulting 1-form $A=A_{0}/\lambda ,\,$%
is homogeneous of degree zero in its coefficient functions, the curvature
features of the implicit hypersurface are completely specified in terms of
the similarity invariants of the Jacobian matrix constructed from the
components of $A.$ \ The curvature similarity invariant of highest degree
(N-1) is defined as the Adjoint curvature, and is equal to the trace of the
Jacobian Adjoint matrix. \ 

It is now well known that given any non-closed 1-form, $A_{0},$of twice
differentiable functions, it is possible to deduce a 2-form of field
intensities, $F_{0},$ and to show that the Maxwell-Faraday equations are
always satisfied. \ However, without additional assumptions, to produce or
define a globally closed charge-current density is another matter. \ An
algorithm from implicit surface theory will be used herein to construct,
from the intrinsic properties of the implicit surface, such a globally
closed N-1 form charge current density. \ The techniques thereby demonstrate
the connection between implicit surface theory and the theory of
electromagnetism. \ Although the Jacobian matrix to be constructed is
globally singular, it is always possible to construct algebraically the
matrix of cofactors transposed, defined as the Jacobian Adjoint matrix. \
Remarkably, multiplication of the covariant components of $A$ by this
singular Adjoint matrix yields an N-1 form density, or current, $J_{s}$,
which is globally closed. \ The result is universally valid if the
coefficients of the 1-form, $A,$ are homogeneous of degree zero in its
component functions. \ The global closure implies that there exists an N-2
form $G_{s}\,$such that $J_{s}-dG_{s}=0.\,$\ The conclusion is valid for any
Holder norm of any signature, of arbitrary isotropic index p, and of
homogeneity index n =1. \ The PDE's associated with this exterior
differential system are known to contain the Maxwell-Ampere equations. \ 

These two Maxwell exterior differential systems lead to another N-1 form
density, previously defined\ \lbrack 1\rbrack\ in the four dimensional case
as Topological Spin, $A\symbol{94}G_{s}.$ \ This N-1 form always satisfies
the equation

\begin{equation}
d(A\symbol{94}G_{s})=F\symbol{94}G_{s}-A\symbol{94}J_{s},
\end{equation}
and demonstrates that twice the difference between the magnetic and electric
energy densities of the field is cohomologous with the interaction energy
density, $A\symbol{94}J_{s}$, generated by interaction of the Potentials and
the Charge-Current density. \ If the Holder norm used to make the initial
1-form homogeneous of degree 1 in its component functions is specialized to
be of euclidean signature, isotropic index p =\ 2, and homogeneity index n
=1 (which defines the Gauss map), then the N-1 form $J_{s}$ is uniquely
determined. \ It is a major result\ of this article to show that interaction
energy density, $A\symbol{94}J_{s},$ is then always proportional to Adjoint
curvature\ of the implicit hypersurface constructed from the 1-form of
Action Potentials. \ On a four dimensional variety, $A\symbol{94}J_{s}$ is
cubic in the principle curvatures. \ 

With respect to a process defined by the induced charge-current density,
Cartan's \ magic formula of topological evolution demonstrates a formal
correspondence to the first law of thermodynamics. The internal energy
density of the physical system described by the 1-form, $A_{0},$ evolving in
the direction field of the closed charge-current density, \ $J_{s}$, is
exactly the coefficient of the Interaction energy N-form, $A\symbol{94}%
J_{s}. $ \ This coefficient is exactly equal to the Adjoint curvature of the
implicit surface. Hence a correspondence is established between the
curvature theory of implicit hypersurfaces, the charge-current density
interaction, and the internal energy of a thermodynamic system. \ \ 

The implicit hypersurfaces can be put into equivalence classes depending
upon the Pfaff dimension or class of the generating 1-form. \ Examples
indicate that, depending upon the Pfaff dimension, the charge current
densities are proportional to the\ adjoint curvatures and/or the topological
torsion induced by the generalized implicit hypersurface.

\ It is important to realize that the method to be discussed involves
curvatures, torsion and energy densities, but does not depend explicitly
upon a metric,\ gauge constraints, or the Einstein field equations. \ In
section 2 some topological and thermodynamic features of electromagnetism
will be discussed. \ In section 3, the theory of generalized implicit
hypersurfaces will be developed. \ In section 4 a number of examples will be
summarized for generalized implicit hypersurfaces in N=3 and N=4 dimensions,
demonstrating the claim that an intrinsic current exists, and that the
intrinsic charge-current interaction with the potentials is equal to the N-1
similarity invariant of the hypersurface. \ The Maple programs that
generated the examples can be downloaded from the internet [9].

\smallskip

\subsection{Some Topological and Geometrical Features of Electromagnetism}

\subsubsection{\protect\vspace{1pt}Charge counting, conductors and insulators%
}

From experience it is known that a given electromagnetic charge-current
density $J$ is conserved: $dJ=0$ (the 4 vector density has zero divergence).
\ However, a more important result is the observation of global charge
neutrality, which can be attributed to a topological idea. \ As $dJ=0,$ is a
global statement, there exists an N-2 form, $G,$ such that $J-dG=0$. \ This
exterior differential system \lbrack 2\rbrack\ is equivalent to the
Maxwell-Ampere system of partial differential equations. \ The integral of $%
G $ over a closed cycle in domains where \ $dG=0\,\ $yields values whose
ratios are rational (Gauss' law of counting charges). \ When the closed
integration domain is a boundary, the net charge enclosed is zero, yielding
charge neutrality. \ These topological aspects can be used to distinguish
insulators from conductors. \ Three dimensional insulators can be separated
in the presence of an external $\mathbf{E}$ field into two physical
components with each component interior enclosed by a two dimensional
boundary. \ The external field distorts the internal charge distribution to
produce a dipole field. \ Each physical component remains charge neutral
when the external field is removed. \ Similarly three dimensional conductors
can be separated into two physical components, but the\ presence of a
remnant exterior electromagnetic field between the components indicates that
the closed two dimensional varieties of each component are cycles, not
necessarily boundaries. \ The components are said to be charged.

\subsubsection{Domains of support}

It has long been respected that physical work is required to produce charge
separation, and that such charge separation leads to a potential difference
between the charged components that can be used to produce useful work. \ In
fact, the conventional physics approach to understanding electromagnetism is
to start with some given distribution of charge currents and compute by some
set of rules the associated potentials. \ In this article, the opposite
procedure is exploited. \ The starting point will be given in terms of a set
of potentials (functions), which can be used to construct a 1-form of
Action, $A_{0},$on a variety of independent variables. \ For C2 functions,
exterior differentiation generates the exterior differential system, $%
F0-dA_{0}=0.$ \ The closure of this system is always equivalent to the
system of PDE's that are known as the Maxwell-Faraday equations. \ The
Maxwell-Faraday exterior differential system indicates that the domain of
support for the 2-form of field intensities is usually open, or compact with
boundary. \ The only possible exceptions are the torus and the Klein bottle.
\ However these exceptions fail if the 2-form is of rank 4.

In this article an algorithm is presented whereby the potentials will lead
to a well defined set of charge-currents, a procedure which is opposite to
the conventional methods. \ However, the new method exploited herein has
geometric and topological significance. \ The 1-form of Action Potentials
will be made homogeneous of degree zero by division by a suitable Holder
norm, $\lambda ,$ leading to the expression, $A=A_{0}/\lambda $. \ The
Jacobian matrix of $A$ will be constructed, as well as its Adjoint (matrix
of cofactors transposed). \ In this sense, the components of 1-form $\ A$
can be interpreted as the normal field to an implicit hypersurface. \ The
similarity invariants of the Jacobian matrix\ determine the important
curvature features of the hypersurface. \ The Jacobian matrix so constructed
will always be singular and often is of maximal rank, N-1. \ The similarity
invariant of highest degree, N-1, is equal to the trace of the Adjoint
Jacobian matrix, and is therefor defined herein as the Adjoint curvature. \
The adjoint curvature plays a dominant part in the discussion that follows.

Multiplication of the components of $A$ by the Adjoint matrix permit the
construction of a closed N-1 form density, which will play the role of a
deduced electromagnetic charge current density, $J_{s}.$ \ The notation
(with a subscript $s$) is such as to remind the reader that this current
density was created from the singular Adjoint matrix generated from the
1-form of Potentials. \ As this N-1 form is closed (has zero divergence
globally) there exists a $G_{s}$ such that $J_{s}-dG_{s}=0.$ \ The PDE's
created by this exterior differential system are equivalent to the
Maxwell-Ampere equations. In a space of 4 dimensions the properties of the
N-2=2 form $G_{s}$ are not the same as the properties of the 2-form $F_{0}.$
\ The domain of support for the 2-form $F_{0}$ is not compact without
boundary, while the domain of support for $G_{s}$ can be compact without
boundary.

\subsubsection{Interaction energy density}

In addition, the interaction energy density, defined as the N-form density, 
\begin{equation}
A\symbol{94}J_{s}=F\symbol{94}G_{s}-d(A\symbol{94}G_{s}),
\end{equation}
will be computed. \ The term $A\symbol{94}G$ in electromagnetic systems has
been previously defined as ''Topological Spin'' [1]\ for it has the physical
dimensions of $\,$joule-s in electromagnetic systems. The term $F\symbol{94}%
G $ represents twice the difference between magnetic and electric energy
density and changes sign from a plasma to an electrostatic state. \ The
equation is a statement of the cohomology of the two forms of energy
density. \ In regions where $A\symbol{94}G$ is closed, the closed 3
dimensional integrals of $A\symbol{94}G$ have values whose ratios are
rational [3]\ and are therefore countable. \ It will be demonstrated below
that this interaction density is precisely equal to the Adjoint curvature of
the hypersurface whose normal field is generated by the 1-form, $A.$ \ On a
variety of four dimensions, this result implies that interaction energy
between the 4 potentials and the deduced (or intrinsic) charge current
density is related to a cubic polynomial of the hypersurface curvatures,
while the Gaussian sectional curvature (and therefor mass energy density)\
is quadratic in the surface curvatures. \ When the Jacobian matrix is of
maximal rank N-2, the interaction energy vanishes. \ Note that if the
interaction energy density is zero, the charge current density need not be
zero. \ A special case exists such that if $J_{s}$ is proportional to the
Topological Torsion 3 form, $A\symbol{94}dA,$ then the interaction energy
density vanishes due to orthogonality of its two components. \ This special
case will be discussed further below.

\subsubsection{Topological evolution and internal energy density}

Given a 1-form of Action $A$ and a closed charge current density $J,$ it is
possible to use Cartan's magic formula [4]\ of topological evolution to
demonstrate a correspondence between the implicit surface theory and the
first law of thermodynamics. \ For evolutionary processes in the direction
of the charge current density, Cartan's magic formula becomes 
\begin{equation}
L(J)A=i(J)dA+d(i(J)A)=W+dU=Q
\end{equation}
Using electromagnetic notation, on a variety $\{x,y,z,t\}$ the (virtual)
work 1-form becomes

\begin{equation}
W=i(J)dA=(\rho \mathbf{E}+\mathbf{J}\times \mathbf{B})_{k}dx^{k}+(\mathbf{J}%
\cdot \mathbf{E})dt
\end{equation}
which is recognized as the product of the Lorentz force density times the
differential displacement plus the dissipative power density times the
increment dt.

In certain cases the induced charge current density, $J_{s}$ will be
proportional to the Topological Torsion field, $A\symbol{94}dA=i(T)dx\symbol{%
94}dy\symbol{94}dz\symbol{94}dt.$ \ (An example of this case is presented
below). \ In such cases, it follows that the evolution of the implicit
surface is given by the expression,

\begin{equation}
L_{(J)}A=L_{(T)}A=i(T)dA+0=(\Gamma )\;A=(\mathbf{E}\cdot \mathbf{B)}A=Q.
\end{equation}
It follows that the heat 3 form and the Topological Torsion 3 form are
proportional:

\begin{equation}
Q\symbol{94}dQ=(\mathbf{E}\cdot \mathbf{B)}^{2}A\symbol{94}dA.
\end{equation}
From classical thermodynamics, when the process produces a heat 1-form $Q$
which does not admit an integrating factor, such a process is
thermodynamically irreversible. \ If the coefficient (the second Poincare
invariant which is related to the 4 form $F\symbol{94}F)$ is not zero, then
irreversible processes exist when the topological torsion of the implicit
surface is not zero. \ In order for $Q$ to admit an integrating factor, the
Frobenius integrability condition must be satisfied, or $Q\symbol{94}dQ=0.$
\ But if the surface 1-form is of Pfaff dimension 4, then $A\symbol{94}%
dA\neq 0$ , and $\mathbf{E}\cdot \mathbf{B}\neq 0.$ \ It follows $Q\symbol{94%
}dQ\neq 0,$ and such irreversible processes are artifacts of 4 dimensions.

Similarly, evaluation of the internal energy density for a process defined
by the dynamics of the charge-current density becomes

\begin{equation}
U=(i(J)A)=\mathbf{A}\cdot \mathbf{J}-\rho \phi .
\end{equation}
which in classical field theory is defined as the interaction energy
density. \ From the discussion above it is apparent that internal energy
density is equivalent to the coefficient of the N form $A\symbol{94}J$, and
at the same time is equal to the Adjoint curvature of the implicit
hypersurface. \ It appears that the charge current interaction energy
density, the thermodynamic internal energy density, and the adjoint
curvature of the implicit surface generated by the 1-form of potentials are
equivalent concepts.

\subsubsection{Gauge constraints are not used.}

It is to be noted that given an N-1 form, $J_{s},$ the N-2 form $G_{s}$ that
satisfies the exterior differential system $J_{s}-dG_{s}=0$ is not unique. \
Any closed N-2 form, $q,\,$such that $dq=0$, may be added to $G_{s}$ without
changing the result of the exterior differential system. \ The same can be
said for the 2-form $F_{0};$ \ there are many closed 1-forms $\gamma ,$ such
that $d\gamma =0$, that can be added to the 1-form $A_{0}$ and yet the
exterior differential system yields the same values for $F_{0}.$ \ It is the
closed integrals of the closed but not exact components of $q$ that
determine the quantized charges. \ It is the closed integrals of closed but
not exact components of $\gamma $ that determine the quantized flux quanta.
\ \ These concepts of ambiguity and non-uniqueness are often parlayed into
specific theories, called gauge theories, where the non-uniqueness is
restricted in form to some equivalence class. \ These problems of specific
gauge equivalence class are not pertinent herein, for the\ results are
formulated to be valid without a specification of gauge.

Classic Maxwell theory, as written in terms of the fields $\mathbf{E}$ and $%
\mathbf{B}$ only, is often said to be a U(1) gauge theory. \ However, when
written in the language of exterior differential forms, Maxwell theory is 
\textit{not} a U(1) gauge theory. \ It is well behaved with respect to the
general linear group, and with respect to differentiable maps without
inverse. \ It is the imposition of zero charge current density (no
interaction energy density) and the Lorentz constitutive constraint that
reduces the general Maxwell theory to a U(1)\ gauge group theory [5]. \ The
usual constraints are such that the both $F\symbol{94}G\,$and $A\symbol{94}J$
vanish simultaneously. \ The constitutive constraints typically imposed are
of the form $\mathbf{B}=\mu \mathbf{H}\,$and $\mathbf{D}=\epsilon \mathbf{E}%
\,\ $with the understanding that $\epsilon \mu c^{2}=1$ and $\mathbf{B\cdot
B-E\cdot E/}c^{2}=0.$ \ An alternate formulation by Bateman\ [6]\ and
Whittaker leads to the constitutive constraint, $\mathbf{D}=\alpha \mathbf{%
B\,}$and $\mathbf{H}=\beta \mathbf{E,}$ with the understanding that $(\alpha
-\beta )$ ($\mathbf{E\cdot B)=0}$, \ The first case can be extended to
include birefringence and Faraday rotation, while the second case
generalizes to include rotational acceleration (Sagnac) effects and Optical
Activity. \ Both simple formulations lead to the result that $F\symbol{94}%
G=0,$ which implies that the N-1 form $A\symbol{94}G\,$\ is closed. \ Hence
(subject to the constraints) over closed N-1 manifolds which are cycles, the
integrals of the Topological Spin N-1=3 form can have ratios which are
rational. \ The closed integrals are evolutionary deformation invariants,
and thereby carry topological information.

\subsubsection{Dissipation}

It is of some importance to note, in certain topological circumstances, that
the Jacobian induced currents, $J_{s},$ occur without the presence of $%
\mathbf{E}$ or $\mathbf{B}$ fields, (when $dA_{0}=F_{0}=0)$ or for other
situations\ where $\mathbf{E}$ or $\mathbf{B}$ fields are present (where $%
F_{0}\neq 0),$ but where the dissipation coefficient $\mathbf{J\circ E}$ is
zero. \ The implicit hypersurface method thereby seems to offer an
alternative, non-quantum mechanical, understanding of what otherwise would
be called superconducting currents. \ In the first case, the $\mathbf{B}$
field is excluded (Meisner effect) from the superconducting region, and in
the second case (Hall effect) a large $\mathbf{B}$ field is present along
with a non-dissipative current. \qquad

If the adjoint curvature of the generalized hypersurface is zero, then
either the charge current density is zero, or the charge current density
resides in the hypersurface, and thereby is orthogonal to the surface
normal. \ In the latter special case, the direction field of the charge
current density is proportional to the Topological Torsion vector [2]\
generated by the 3-form $A\symbol{94}dA$. \ \ Examples below indicate that
there is a correlation between non-zero adjoint curvature and/or topological
torsion and the existence of a charge current density.

\section{Implicit Hypersurfaces}

The algebra of the processes to be described can be staggering, especially
for older folks who do not see too well. \ Hence a Maple program [9] has
been provided to make the computation of examples a bit faster.

\subsection{\protect\vspace{1pt}The normal and tangent fields}

The classic implicit surface is generated by assigning a constant value to a
function, $\phi (x,y,z..).$ \ It is important to recall that an implicit
surface, in contrast to a parametric surface, can consist of more than one
disconnected components. \ The gradient field to the given function
represents a normal field to the surface, and tangent vectors which reside
on the surface are orthogonal to the normal field at all points. \ \ As the
normal field for the classic implicit surface is a gradient field, its
associated 1-form is exact. \ If this normal gradient field is rescaled by a
factor such that is homogeneous of degree zero in its functional arguments,
then the Jacobian matrix of the rescaled normal field can be used to
generate the curvatures of the implicit surface.

This procedure can be extended to the study of \ generalized implicit
surfaces whose normal field is not representable by an exact 1-form. The
1-form representing the normal field can have arbitrary Pfaff dimension. \
If the Pfaff dimension of the 1-form is greater than 2, then the implicit
surface can support topological torsion, $A\symbol{94}dA\neq 0.$ \ It is
necessary that the Pfaff dimension be greater than 2 if the implicit surface
admits an envelope. \ A N-1 tangent vector basis can be constructed
algebraically from the 1-form that represents the normal field. \ In the
language of exterior differential forms, the tangent vectors, $\mathbf{e}$,
have been described as the associate vectors relative to the 1-form $A_{0}$,
and satisfy the equation, $i(\mathbf{e})A_{0}=0.$ \ The array of tangent
vectors and the normal field can be used to form a ND basis at any point in
the implicit surface.

\ In fact it follows that starting from an arbitrary 1-form, defined herein
as the Action 1-form of Potentials, $A_{0},$\ on a variety of independent
variables $(x,y,z,...),$ it is possible to develop the curvature properties
of the generalized implicit surface algebraically after admitting only one
differentiation process. \ The components of $A_{0}$ will play the role of
the normal field. \ 

\smallskip

\subsection{The homogeneous Holder norm and similarity curvature invariants.}

After division by a suitable function of the coefficient potentials, $%
\lambda ,\,$the original 1-form of Action 
\begin{equation}
A_{0}=(U(x,y,z,...)dx+V(x,y,z,...)dy+W(x,y,z,...)dz...),
\end{equation}
can be made homogeneous of degree zero in terms of those coefficient
functions that define the potentials. \ It is to be emphasized that the
homogeneity condition is not on the arguments of the coefficients, but on
the coefficient functions themselves. \ \ The scaling function of choice, $%
\lambda ,\,$is a Holder norm and is defined in terms of the covariant
coefficients of the 1-form: 
\begin{equation}
\lambda =(aU^{p}+bV^{p}+cW^{p}+...)^{n/p}.\vspace{1pt}
\end{equation}
The index $n$ will be defined as the homogeneity index; the index $p$ will
be described herein as the isotropic index, and the constants $(a,b,c...)$
are constant scale factors whose signs determine the signature. \ By
choosing the index $n$ to be unity, $n=1$, the 1-form, $A$, defined as

\begin{equation}
A=A_{0}/\lambda =(Udx+Vdy+Wdz...)/\lambda =A_{k}dx^{k}
\end{equation}
becomes homogeneous of degree zero in its coefficients. \ That is if every
coefficient function is increased by a factor $\beta $ then the coefficient
function $A_{k}$ does not change. \ \ This homogeneous degree zero 1-form, $%
A_{0}/\lambda ,$ is used to define an implicit hypersurface in the variety,
whose geometrical properties can be expressed classically in terms of the
similarity invariants of the associated \textit{singular} Jacobian dyadic
(or matrix). \ Classically these similarity invariants are ''symmetric''
functions of the surface curvatures. \ Examples are given below.

The doubly covariant Jacobian dyadic of coefficients is defined as the
matrix of functions

\begin{equation}
Jacobian(A)=\left[ \partial A_{m}/\partial x^{n}\right] =\lbrack
JAC(A)\rbrack _{mn}=\left[ \Bbb{J}\right]
\end{equation}
The determinant of the Jacobian matrix so constructed ($n=1$, any $%
a,b,c...,p $) ) is always zero, indicating the existence of at least one
zero eigen value (curvature or reciprocal radius). \ Hence the Jacobian
matrix so constructed is singular, and induces a singular metric on the
variety via the pullback $\lbrack g\rbrack \,=\lbrack \Bbb{J}\rbrack
^{^{Transpose}}\circ \lbrack \Bbb{J}\rbrack .$ \ \ The zero determinant
result also implies the existence of a global N-1 dimensional variety which
in effect defines the implicit (hyper)\ surface. \ It is a standard
geometrical procedure to construct the symmetric similarity invariants of
the Jacobian matrix by forming the Cayley-Hamilton characteristic
polynomial. \ Note that the induced symmetric metric $\lbrack g\rbrack $
does not carry the complete story of the surface properties inherent in the
Jacobian dyadic, for the Jacobian matrix is not necessarily symmetric. \ As
pointed out by Brand \lbrack 7\rbrack , the anti-symmetric components of the
Jacobian dyadic also have important invariance properties. \ These
additional invariants are developed in terms of the 2-form $F=dA.$

\smallskip

\subsection{A globally closed current from the adjoint matrix}

Next construct the doubly contravariant matrix $\left[ \widehat{\Bbb{J}}%
\right] $ equal to the adjoint (matrix of co-factors transposed) of the
doubly covariant Jacobian matrix. \ This adjoint matrix exists
algebraically, even though the inverse of the singular Jacobian matrix, and
the inverse of the induced singular metric does not. \ Use the adjoint
matrix to construct the contravariant vector current, $\left| \mathbf{J}%
_{s}\right\rangle ,$

\begin{equation}
\left| \mathbf{J}_{s}\right\rangle =[ADJ(A)]^{nm}\cdot \left| \mathbf{A}%
\right\rangle =[\widehat{\Bbb{J}}]^{nm}\cdot \left| \mathbf{A}\right\rangle ,
\end{equation}
and the N-1 form density, $J_{s}:$

\begin{equation}
J_{s}=i(\mathbf{J}_{s})dx\symbol{94}dy\symbol{94}dz...
\end{equation}
Remarkably for any Holder norm with $n=1$, arbitrary signature, arbitrary
scale factors, and arbitrary exponent p, the N-1 form, $J_{s}$, is closed.

\begin{equation}
dJ_{s}=0\,\,\,\,\,\,\,for\,\,\,n=1
\end{equation}
As this closure result is global, it follows that $J_{s}-dG_{s}=0,.$ which
is equivalent to the Maxwell-Ampere equations. \ The subscript $s$ is used
to distinguish the fact that $J_{s}$ has been deduced from the singular
Jacobian matrix, and does not explicitly depend upon the field intensities, $%
F_{0}=dA_{0}$ and some arbitrary constitutive constraint between $F$ and $G.$
\ Note that given a $J_{s}$ the corresponding $G_{s}$ is not uniquely
determined. \ The N-2 form density, $G_{s},$ may have closed and exact
components as well as closed non-exact components, neither of which
contribute to a specific Charge-Current.

\vspace{1pt}It is to be noted the induced metric\ is singular and therefor
cannot be used to define a raising tensor as an inverse metric. \ Yet a
raising tensor field $[\widehat{\Bbb{J}}]^{nm}$ can be functionally well
defined in terms of the Adjoint of the Jacobian matrix. \ This raising
tensor field, unlike a non-singular metric inverse field, is not\ symmetric.
\ \ Moreover the Adjoint method applies to 1-forms (and therefor
hypersurfaces) that do not satisfy the Frobenius condition of unique
integrability. \ Hence, topological torsion, defined as the 3-form, $A_{0}%
\symbol{94}dA_{0},$ need not be zero. \ It will be demonstrated below that
when the Holder norm is specialized to the Gauss map, $a=b=c=...=1,\,\,p=2,%
\,n=1,$ then the coefficient of the interaction N form density, $A\symbol{94}%
J_{s},$ is equal to the ($N-1)^{th}$ similarity invariant of the Jacobian
field. \ For all implicit surfaces, simple or not, this similarity invariant
is equal to the trace of the Jacobian Adjoint matrix and is equal to the sum
of all possible products of degree N-1 of the eigen values of the Jacobian
matrix. \ This similarity invariant will be defined as the Adjoint Curvature
of the implicit surface. \ In 3 dimensions the Adjoint curvature\ of simple
implicit surfaces is equal to the Gauss sectional curvature. \ 

\smallskip

\subsection{The adjoint curvature and the interaction energy}

In summary, a well defined procedure has been implemented to deduce a
consistent exterior differential system in Maxwell - Electromagnetic format,
starting from a set of potentials that define the coefficients of a 1-form
of Action, $A_{0}.$ \ It follows that the exterior differential system $%
F-dA=0$ is always equivalent to the system of PDE's known as the
Maxwell-Faraday equations. \ \ The induced system described above, $%
J_{s}-dG_{s}=0,$ generates the system of PDE's known as the Maxwell-Ampere
equations. \ Note that no constitutive or duality constraints have been
subsumed. \ It is known that the two combined exterior differential systems
lead to a N-1 form, previously defined as topological spin, $A\symbol{94}%
G,\,\,\,$and a third exterior differential system, $d(A\symbol{94}G)-F%
\symbol{94}G+A\symbol{94}J=0.$ \ \ The last term, defined as the interaction
energy and equal to $A\symbol{94}J_{s},$ can be evaluated in terms of the
curvature invariants of the implicit hypersurface generated by the 1-form of
Action Potentials. \ It is remarkable that the term $A\symbol{94}J_{s}$ is
equal to the volume element multiplied by the ($N-1)^{th}$ similarity
invariant, defined as the $Trace\lbrack \widehat{\Bbb{J}}\rbrack .$

\begin{equation}
Interaction\_energy\,=\,A\symbol{94}J_{s}=Trace[\widehat{\Bbb{J}}]dx\symbol{%
94}dy\symbol{94}dz...
\end{equation}
As the ($N-1)^{th}$ similarity invariant can be interpreted in terms of a
polynomial cubic in the curvatures of the hypersurface in 4D, it would
appear that concept of the interaction energy between the charge current
density and the potentials can be related to an expression cubic in the
curvatures of the associated hypersurface.

\section{\protect\bigskip Examples}

The following examples will display some of the features of\ the theory of
generalized implicit hypersurfaces in 3 and 4 dimensions. The 3D examples
can be of two physically interesting categories based on the coordinate sets 
$\{x,y,z\}$ and $\{x,y,t\}$ \ From an electromagnetic interpretation, the
first category has the properties of a 3D plasma. \ The second category
admits an $\mathbf{E}$ field as well as a $\mathbf{B}$ field. \ Each of
these\ 3D categories can be viewed as special cases of the 4D\ category
based on coordinate variables of the type $\{x,y,z,t\}$ \ In addition there
is a topological refinement of the categories which depend upon the Pfaff
dimension of the 1-form, $A_{0},$ used to model the normal direction field
of the implicit hypersurface. \ Most classical developments of implicit
simple surface theory study those cases where the Pfaff dimension of the
1-form,.$A$, is unity. \ Such spaces do not support topological torsion. \ A
rotating spherical surface does not support torsion. \ An expanding
spherical surface does not support torsion. \ However an expanding and
rotating spherical surface does support topological torsion. \ 

Make sure you are aware that symbolic math programs using Maple programs [9]
have been provided for you to check the details and extend the examples
presented below.

\subsection{Simple surfaces\ of one component in 3D.}

For simplicity, consider those surfaces generated by fixed values assigned
to functions of the form $\phi =f(x,y)-z.$ \ Such surfaces are of a single
component and do not support a non-zero 2-form \ The differential of the
function $\phi $ generates the exact 1-form

\begin{equation}
A_{0}=(\partial \phi /\partial x)dx+(\partial \phi /\partial y)dy+(\partial
\phi /\partial z)dz=(\partial f/\partial x)dx+(\partial f/\partial y)dy-dz
\end{equation}
The 1-form associated with such surfaces is of Pfaff dimension 1. \ \ Choose
the Holder norm equivalent to the Gauss map

\begin{equation}
\lambda =\{(\partial f/\partial x)^{2}+(\partial f/\partial y)^{2}+1\}^{1/2}.%
\vspace{1pt}
\end{equation}
and construct the homogenous of degree zero 1-form

\begin{equation}
A=A_{0}/\lambda .
\end{equation}
The homogeneous 1-form, $A$, can be of Pfaff dimension 2. \ Form the
Jacobian matrix, $\left[ \partial A_{m}/\partial x^{n}\right] =\left[ \Bbb{J}%
\right] ,$ of the covariant components of the 1-form, $A,$and construct the
similarity invariants, and the induced current. \ For this simple surface it
is assumed that $\phi $ is linear in $z$. \ The determinant of the Jacobian
matrix vanishes, which implies that the Jacobian matrix is singular and has
no inverse. \ The remaining similarity invariants are:

\begin{eqnarray}
Mean\_Curvature &=&-1/2\{(\partial ^{2}f/\partial x^{2})(1+(\partial
f/\partial y)^{2})+(\partial ^{2}f/\partial y^{2})(1+(\partial f/\partial
x)^{2})  \nonumber \\
&&-2(\partial f/\partial x)(\partial f/\partial y)\partial ^{2}f/\partial
x\partial y\}/\lambda ^{3}
\end{eqnarray}
and

\begin{eqnarray}
Adjoint\_Gauss\_Curvature &=&Trace[\widehat{\Bbb{J}}]  \nonumber \\
&=&\{(\partial ^{2}f/\partial x^{2})(\partial ^{2}f/\partial
y^{2})-(\partial ^{2}f/\partial x\partial y)^{2}\}/\lambda ^{4}
\end{eqnarray}
The induced current is of the form, $[0,0,J_{z}]$ where

\begin{equation}
J_{z}=\{(\partial ^{2}f/\partial x^{2})(\partial ^{2}f/\partial
y^{2})-(\partial ^{2}f/\partial x\partial y)^{2}\}/\lambda ^{3}.
\end{equation}
It follows that the N-form $A\symbol{94}J$ becomes

\begin{eqnarray}
A\symbol{94}J_{s} &=&(Adjoint\_Gauss\_Curvature)\,dx\symbol{94}dy\symbol{94}%
dz  \nonumber \\
&=&\{(\partial ^{2}f/\partial x^{2})(\partial ^{2}f/\partial
y^{2})-(\partial ^{2}f/\partial x\partial y)^{2}\}dx\symbol{94}dy\symbol{94}%
dz/\lambda ^{4}
\end{eqnarray}
and the coefficient of the interaction is precisely equal to the Adjoint
curvature, which is equivalent to the classic Gauss curvature of the
implicit surface in the 3 dimensional variety.

For exact 1-forms, the 2-form $F_{0}=dA_{0}$ vanishes. \ Hence the general
formula

\begin{equation}
A_{0}\symbol{94}J_{s}=F_{0}\symbol{94}G_{s}-d(A_{0}\symbol{94}G_{s})
\end{equation}
becomes

\begin{equation}
A\symbol{94}J_{s}=-d(A_{0}\symbol{94}G_{s})/\lambda .
\end{equation}
This result is equivalent to the Chern statement of the Gauss-Bonnet
theorem: the Gauss curvature is integrable [9].

The singular induced (pullback) metric is given by the expression,

\begin{equation}
\left[ g\right] =\left[ 
\begin{array}{lll}
(\partial ^{2}f/\partial x^{2})^{2}+(\partial ^{2}f/\partial x\partial y)^{2}
& (\partial ^{2}f/\partial x\partial y)(\partial ^{2}f/\partial
x^{2}+\partial ^{2}f/\partial y^{2}) & 0 \\ 
(\partial ^{2}f/\partial x\partial y)(\partial ^{2}f/\partial x^{2}+\partial
^{2}\phi /\partial y^{2}) & (\partial ^{2}f/\partial y^{2})^{2}+(\partial
^{2}f/\partial x\partial y)^{2} & 0 \\ 
0 & 0 & 0
\end{array}
\right] /\lambda ^{2},
\end{equation}
and can be used to construct a line element, $(ds)^{2}$ on the two
dimensional subspace.\smallskip

\subsection{Classical Implicit Surfaces with more than one component in 3D.}

The classic implicit surface is defined by fixed values assigned to
non-linear functions of the form $\phi (x,y,z).$ \ The differential of the
function $\phi $ generates the exact 1-form of Pfaff dimension 1:

\begin{equation}
A_{0}=(\partial \phi /\partial x)dx+(\partial \phi /\partial y)dy+(\partial
\phi /\partial z)dz.
\end{equation}
As before, choose the Holder norm equivalent to the Gauss map

\begin{equation}
\lambda =\{(\partial \phi /\partial x)^{2}+(\partial \phi /\partial
y)^{2}+(\partial \phi /\partial z)^{2}\}^{1/2}.\vspace{1pt}
\end{equation}
and construct the homogenous of degree zero 1-form (which can be of Pfaff
dimension 2):

\begin{equation}
A=A_{0}/\lambda .
\end{equation}
Create the Jacobian matrix, $\left[ \partial A_{m}/\partial x^{n}\right] ,$
of the covariant components of the 1-form, $A,$and construct the similarity
invariants, and the induced current. Note that the Jacobian matrix is
symmetric. \ The determinant of the Jacobian vanishes indicating the matrix
is singular and without inverse. \ Then the remaining similarity invariants
are constructed from the trace of the Jacobian matrix and the trace of the
Adjoint matrix. \ The curvature formulas are best computed via a symbolic
math program such as Maple \lbrack 9\rbrack .

\ The mean curvature becomes\vspace{1pt} 
\begin{eqnarray}
&&Mean\;Curvature \\
&=&-\{2(\partial \phi /\partial x)(\partial \phi /\partial y)(\partial
^{2}\phi /\partial x\partial y)-\partial ^{2}\phi /\partial z^{2}\left(
(\partial \phi /\partial y)^{2}+(\partial \phi /\partial x)^{2}\right) 
\nonumber \\
&&+2(\partial \phi /\partial y)(\partial \phi /\partial z)(\partial ^{2}\phi
/\partial z\partial y)-\partial ^{2}\phi /\partial x^{2}\left( (\partial
\phi /\partial y)^{2}+(\partial \phi /\partial z)^{2}\right)  \nonumber \\
&&+2(\partial \phi /\partial z)(\partial \phi /\partial x)(\partial ^{2}\phi
/\partial x\partial z)-\partial ^{2}\phi /\partial y^{2}\left( (\partial
\phi /\partial x)^{2}+(\partial \phi /\partial z)^{2}\right) \}/3\lambda ^{3}
\nonumber
\end{eqnarray}

and the Adjoint - Gauss curvature becomes

\begin{eqnarray}
&&Adjoint-Gauss\;Curvature \\
&=&-\{2(\partial \phi /\partial x)(\partial \phi /\partial y)(\partial
^{2}\phi /\partial x\partial y)(\partial ^{2}\phi /\partial z^{2})-(\partial
^{2}\phi /\partial y^{2})(\partial ^{2}\phi /\partial x^{2})(\partial \phi
/\partial z)^{2}  \nonumber \\
&&+2(\partial \phi /\partial y)(\partial \phi /\partial z)(\partial ^{2}\phi
/\partial z\partial y)(\partial ^{2}\phi /\partial x^{2})-(\partial ^{2}\phi
/\partial z^{2})(\partial ^{2}\phi /\partial y^{2})(\partial \phi /\partial
x)^{2}  \nonumber \\
&&+2(\partial \phi /\partial z)(\partial \phi /\partial x)(\partial ^{2}\phi
/\partial x\partial z)(\partial ^{2}\phi /\partial y^{2})-(\partial ^{2}\phi
/\partial x^{2})(\partial ^{2}\phi /\partial z^{2})(\partial \phi /\partial
y)^{2}  \nonumber \\
&&+(\partial \phi /\partial x)^{2}(\partial ^{2}\phi /\partial y\partial
z)^{2}+(\partial \phi /\partial y)^{2}(\partial ^{2}\phi /\partial z\partial
x)^{2}+(\partial \phi /\partial z)^{2}(\partial ^{2}\phi /\partial x\partial
y)^{2}  \nonumber \\
&&-2(\partial \phi /\partial x)(\partial \phi /\partial y)(\partial ^{2}\phi
/\partial z\partial x)(\partial ^{2}\phi /\partial z\partial y)  \nonumber \\
&&-2(\partial \phi /\partial y)(\partial \phi /\partial z)(\partial ^{2}\phi
/\partial x\partial y)(\partial ^{2}\phi /\partial x\partial z)  \nonumber \\
&&-2(\partial \phi /\partial z)(\partial \phi /\partial x)(\partial ^{2}\phi
/\partial x\partial y)(\partial ^{2}\phi /\partial x\partial z)\}/\lambda
^{4}  \nonumber
\end{eqnarray}

\vspace{1pt}The induced current $J_{s}$ may be computed by multiplying the
components of $A$ with the Adjoint matrix relative to the Jacobian matrix,
and it may be shown that the interaction N form is precisely equal to the
volume element multiplied by the trace of the Adjoint matrix.

\begin{equation}
A\symbol{94}J_{s}=(Adjoint\_Gauss\_Curvature)\,dx\symbol{94}dy\symbol{94}dz
\end{equation}

\subsection{\protect\smallskip Implicit surfaces of the Bateman type}

A generalized implicit surface generated by a 1-form which is of Pfaff
dimension 2 has a representation of the Bateman type. \ That is, $%
A_{0}=\alpha (x,y,z)db(x,y,z).$ \ The procedures are the same as above. \
Choose a Holder norm in the form of a Gauss map such that

\begin{equation}
\lambda =\alpha \{(\partial \beta /\partial x)^{2}+(\partial \beta /\partial
y)^{2}+(\partial \beta /\partial z)^{2}\}^{1/2}.\vspace{1pt}
\end{equation}
Compute the Jacobian matrix which now can have an anti-symmetric part

\subsection{\ Non integrable 2-surfaces in 3D}

The procedure is the same as above, except now the 1-form \ $A_{0}$\ has
arbitrary coefficients,

\begin{equation}
A_{0}=U(x,y,z)dx+V(x,y,z)dy+W(x,y,z)dz=
\end{equation}
As before, choose the Holder norm equivalent to the Gauss map

\begin{equation}
\lambda =\{(U)^{2}+(V)^{2}+(W)^{2}\}^{1/2}.\vspace{1pt}
\end{equation}
and construct the homogenous of degree zero 1-form

\begin{equation}
A=A_{0}/\lambda .
\end{equation}
Form the Jacobian matrix, $\left[ \partial A_{m}/\partial x^{n}\right] ,$ of
the covariant components of the 1-form, $A,$and construct the similarity
invariants, and the induced current. \ The results are the same for the
interaction N-form $A\symbol{94}J.$:

\begin{equation}
A\symbol{94}J_{s}=(Adjoint\_Curvature)\,dx\symbol{94}dy\symbol{94}dz
\end{equation}

It should be remarked that the same procedures are valid in dimension N. \
If \ $\Omega \,\,$is the N dimensional differential volume element, then in
N dimensions,

\begin{equation}
A\symbol{94}J_{s}=(Adjoint\_Curvature)\,\,\Omega
\end{equation}
The computations can be lengthy, and it is advised that the symbolic math
program provided be used [9].

\smallskip

\subsection{An interpretation in terms of Electromagnetism}

\subsubsection{The 3-D Variety is Spatial: (x,y,z)}

In this example, the variety of independent variables $\{x,y,z\},$ is
presumed to be independent of time. \ Given the 1-form $A_{0}=Udx+Vdy+Wdz,$
it is possible to construct the field intensities, $F_{0}=dA_{0}\,$on the
presumption that the 1-form is the 1-form of Action potentials for an
electromagnetic field. \ By construction there is no time dependence and no
scalar potential on the 3 D spatial domain. \ The 2-form $F\,$\ of field
intensities consists of 3 components related to the curl of the vector
potential, $\mathbf{A}_{0}=[U,V,W]:$ 
\begin{equation}
\mathbf{B}_{0}=curl\,\mathbf{A}_{0}\mathbf{.}
\end{equation}
By computing the Jacobian matrix of the rescaled 1-form, $A,$ which is
homogeneous of degree 0, the above procedures lead to a divergence free
current. \ This current is a globally closed N-1=2-form, and so is related
to the exterior derivative of some N-2=1-form. \ In this example, the 1-form
of field excitations is defined by the symbolism, 
\begin{equation}
G=H^{x}dx+H^{y}dy+H^{z}dz=\mathbf{H\cdot dr.}
\end{equation}
\ It follows that the induced current is of the form 
\begin{equation}
\mathbf{J}_{s}=curl\,\mathbf{H,}
\end{equation}
but although there is a current there is no analogue to a charge density
distribution for the time independent 3 dimensional format.

The interaction N=3-form becomes

\begin{eqnarray}
A\symbol{94}J_{s} &=&(Adjoint\_Gauss\_Curvature)\,dx\symbol{94}dy\symbol{94}%
dz  \nonumber \\
&=&(\mathbf{A\cdot J}_{s})dx\symbol{94}dy\symbol{94}dz
\end{eqnarray}
The topological Spin N--1=2 form becomes

\begin{equation}
A_{0}\symbol{94}G=i(\mathbf{A}_{0}\mathbf{\times H})dx\symbol{94}dy\symbol{94%
}dz
\end{equation}
and has a divergence equal to the first ''Poincare'' invariant,

\begin{equation}
d(A_{0}\symbol{94}G)=\{(\mathbf{B}_{0}\mathbf{\cdot H)}-(\mathbf{A}_{0}%
\mathbf{\cdot J}_{s})\}dx\symbol{94}dy\symbol{94}dz.
\end{equation}
(The symbol $i(\mathbf{A}_{0}\mathbf{\times H})$ is used for interior
product operator and should not be confused with the imaginary $\sqrt{-1}$).
\ The result is remarkable for it implies that the closed integral of the
magnetic energy density $(\mathbf{B}_{0}\mathbf{\cdot H)}$ minus the
Gaussian curvature times $\lambda ,$\thinspace $(or\,\,\mathbf{A}_{0}\mathbf{%
\cdot J}_{s})$ of the surface created by the non-exact 1-form, $A_{0}$, is
an invariant of any steady flow process on the variety $\{x,y,z\}.\ $

Note that it is not apparent nor true that $\mathbf{B}_{0}$ is linearly
related to $\mathbf{H}$, where curl $\mathbf{H}$ is the source of the
Adjoint closed current. \ As the 1-form is not necessarily integrable, the
helicity $\mathbf{A}_{0}\mathbf{\cdot B}_{0}$ is not necessarily zero. \ \
In hydrodynamics, the spatial parts of the Action 1-form can be related to
the fluid dynamics velocity field, and the $\mathbf{B}_{0}$ field is the
fluid vorticity. \ In the integrable case the two direction fields
associated with $\mathbf{A}_{0}$ and $\mathbf{B}_{0}$ form a surface (the
Lamb surface).

Note that for the non-exact hypersurfaces in 3 dimensions, the similarity
invariant is no longer a perfect differential, but is modified by the
presence of the enstrophy (sqaure of the vorticity or $\mathbf{B}_{0}$
field). \ This example and others are displayed in [9].

\subsubsection{The 3D Variety is 2+1 time dependent (x,y,t)}

\vspace{1pt}In this example, the variety of independent variables $%
\{x,y,t\}, $ is presumed to consist of two spatial variables and time. \
Given the 1-form $A_{0}=Udx+VdY-\phi dt,$ it is possible to construct the
field intensities, $F_{0}=dA_{0}\,$on the presumption that the 1-form is the
1-form of Action potentials for an electromagnetic field. \ Each of the
component functions can be functions of \ $\{x,y,t\}$ and $\phi (x,y,t)\;$%
will play the role of the scalar potential. The 2-form of field intensities, 
$F\,$consists of one magnetic component, orthogonal to the xy plane (which
means that it is in the $t$ direction), and two electric components. \ In
engineering format: 
\begin{equation}
\mathbf{B}_{t}=\partial V/\partial x-\partial U/\partial
y\,\,\,\,\,\,\,\,\,\,\,\mathbf{E}_{x}=-\partial U/\partial t-\partial \phi
/\partial x\,\,\,\,\,\,\ \,\mathbf{E}_{y}=-\partial V/\partial t-\partial
\phi /\partial y
\end{equation}
By computing the Jacobian matrix of the rescaled 1-form which is homogeneous
of degree 0, the above procedures lead to a divergence free current. \ This
divergence free current is\ a globally closed N-1=2-form, and so is related
to the exterior derivative of some N-2=1-form. \ In this example, the 1-form
of field excitations is defined in terms of the symbols, 
\begin{equation}
G_{s}=D^{y}dx-D^{x}dy+H^{t}dt\mathbf{.}
\end{equation}
\ It follows that, in engineering notation, the induced close current has
the classic format, 
\begin{eqnarray}
\mathbf{J}_{s} &=&[\partial \,H^{t}/\partial y+\partial \,D^{x}/\partial
t,\,\partial \,H^{t}/\partial x+\partial \,D^{y}/\partial t,+\partial
D^{x}/\partial x+\partial D^{x}/\partial x] \\
&=&[curl\,\mathbf{H}^{t}+\partial \,\mathbf{D}/\partial t]  \nonumber \\
\rho &=&div\mathbf{D.}
\end{eqnarray}
In contrast to the previous example, it is now apparent that this time
dependent system can have a non-zero charge distribution as well as a
current. \ Note that the format for $G_{s}$ is the canonical form of the
Heisenberg system.

The interaction N=3-form becomes as before,

\begin{eqnarray}
A\symbol{94}J_{s} &=&(Adjoint\_Gauss\_Curvature)\,dx\symbol{94}dy\symbol{94}%
dz  \nonumber \\
&=&(\mathbf{A\cdot J}_{s}-\rho \phi )dx\symbol{94}dy\symbol{94}dz.
\end{eqnarray}
The topological Spin N--1=2 form becomes in component form,\bigskip

\begin{equation}
A_{0}\symbol{94}G_{s}\Rightarrow \lbrack \mathbf{A}_{0}H^{t}+\mathbf{D}\phi ,%
\mathbf{A}_{0}\mathbf{\cdot D}],
\end{equation}
and has a divergence equal to the first ''Poincare'' invariant,

\begin{equation}
d(A_{0}\symbol{94}G_{s})=\{(B_{t}\cdot H^{t}-\mathbf{D\cdot E)}-(\mathbf{A}%
_{0}\mathbf{\cdot J}_{s}-\rho \phi _{0})\}dx\symbol{94}dy\symbol{94}dz.
\end{equation}
The result is remarkable for it implies that the closed integral of the
Lagrangian energy density $(B_{t}\cdot H^{t}-\mathbf{D\cdot E)}$ minus the
Gaussian curvature times $\lambda ,$\thinspace $(or\,\,\mathbf{A}_{0}\mathbf{%
\cdot J}_{s}-\rho \phi _{0})$ of the surface created by the non-exact
1-form, $A_{0}$, is an invariant of any flow process on the variety $%
\{x,y,t\}.\ \ \ $For details see [9].

\smallskip

\subsection{Four Dimension Hypersurfaces}

In four dimensions, the analysis above continues to be true, with the
fundamental result that the interaction density is related to the Adjoint
curvature of the hypersurface defined by the 1-form of Action with
coefficients that are homogeneous of degree zero. \ However, in 4
dimensions, it is possible to distinguish between the mean curvature, $Mean$%
, the Gauss sectional curvature, $Gauss$, and the Adjoint curvature, $Kubic$%
. \ The set $\{H,M,K\}$ are known as the similarity invariants of the
Jacobian matrix. \ The Mean curvature, $Mean$, is proportional to the sum of
the eigen values. \ The Sectional Gauss curvature $Gauss$ is related to the
sum of the three paired products of the eigenvalues of the Jacobian matrix \
For the 1-form which has been made homogeneous of degree zero in its
component functions, the Adjoint curvature $Kubic$ is the unique product of
the three non-zero eigen values of the Jacobian matrix. \ . \ Again the
interaction energy density N=4 form, $A\symbol{94}J_{s}$ has a coefficient
exactly equal to the Adjoint similarity invariant, if the Holder norm is
equivalent to the Gauss map with isotropic index p=2 and homogeneity index n
= 1. \ 

This observation yields a remarkable difference between mass\ energy density
(related to Gaussian sectional second order curvature) and interaction
energy density between the charge current density and the potentials
(related to the Adjoint third order curvature). \ There are three situations
of interest: Case 1, the charge-current density is zero, so that $J_{s}=0$
and $A\symbol{94}J_{s}=0\,$; Case 2, the charge-current density is not zero,
so that $J_{s}\neq 0$ and $A\symbol{94}J_{s}=0$ because of orthogonality;
Case 3, the charge current density is not zero and not orthogonal to 1-form
of Action, such that $A\symbol{94}J_{s}\neq 0$, and $J_{s}\neq 0.$ \ On a
space of N = 4 dimensions there are 3 direction fields that are orthogonal
to the one form of potentials. If the 1-form $A$ is of Pfaff dimension 4,
then there is a unique vector direction field $V$ such that $i(V)dA=\Gamma
A, $ and yet $V$ is orthogonal to $A.$ \ This direction field is determined
by the topological torsion vector $A\symbol{94}dA,$with $\Gamma $
proportional to the coefficient of the 4 form of topological parity, $dA%
\symbol{94}dA.$ $\ $When $\Gamma \Rightarrow 0$, the Pfaff dimension is
three, and the direction field generated by the topological torsion 3-form, $%
A\symbol{94}dA, $ becomes a characteristic vector field of the 1-form $A.$ \
Characteristic vector fields are homeomorphisms that preserve topology.

Examples indicate that Case 1, $J_{s}=0,$ is satisfied if the scalar
potential is zero (any vector potential) or if the vector potential is zero
(any scalar potential), for then the charge current density does not exist
and the Adjoint curvature vanishes. The case which admits only a Scalar
potential yields a system with $\mathbf{E}$ fields and no $\mathbf{B}$
fields. \ A time independent Vector potential without \ a scalar potential
yields the opposite situation with $\mathbf{B}$ fields and no $\mathbf{E}$
fields. \ The Time dependent vector field case admits both $\mathbf{B}$
fields and $\mathbf{E}$ fields.

Examples indicate that Case 3 is usually satisfied if the 1-form of
potentials has both a scalar and vector components which are both time and
space dependent. \ A special situation occurs if the potentials are not
explicitly time dependent, for then the spatial current density is zero but
the charge density is not zero. \ The Adjoint curvature is not zero, and the
interaction energy density does not vanish. \ This example gives credence to
the suggestion that the origin of charge density is cubic curvature. \ The
time independent case can support non zero topological Torsion and non-zero
topological Parity.

In the examples presented below, it appears that if a charge current density
exists, and it does not describe an irreversible process, then it is related
to the third order curvature invariants of the implicit surface defined by
the 1-form of Potentials. \ The algebra is a bit formidable. \ Hence a Maple
program is presented [9]\ for which the reader can verify the computations,
and modify the program to check his own hierarchy of examples. Only the
results of the computations are described here. \ Only the isotropic Holder
norm equivalent to the Gauss map, where p = 2, n =1 and with euclidean
signature, is utilized in the examples presented below unless specified
otherwise

\subsubsection{4D Example 1, Time dependent Scalar potential only}

When applied to a 1-form that consists of a single (time and spatially
dependent) scalar potential, \ 
\begin{equation}
A_{0}=-\phi (x,y,z,t)\,dt,
\end{equation}
it follows that the implicit surface is flat. \ All of the curvature
similarity invariants vanish. \ \ First, renormalize $A_{0}\,$by dividing
through by a Holder norm, $\lambda ,\,$such as to make the new 1-form
homogeneous of degree 0. $\ $Then construct the Jacobian matrix of the
components of $A=A_{0}/\lambda $. \ It follows for the example that all of
the curvature similarity invariant vanish.

\begin{eqnarray}
Mean &=&0, \\
Gauss &=&0 \\
Kubic &=&0.
\end{eqnarray}
\ There is zero induced charge current density;.$J_{s}=0.$ There can be an
electric field, $\mathbf{E}$, but no magnetic field, $\mathbf{B}$. \ The
interaction energy as well as the\ Adjoint curvature are zero; $A\symbol{94}%
J_{s}=0$. $\ $The Pfaff dimension of the 1-form is at most 2. \ The
hypersurface may be considered to be a 3-plane orthogonal to the time
coordinate. \ If the scalar potential is independent from time, the problem
is related to classical electrostatics. \ The implicit surface is flat,
without bending or tension. \ The Coulomb potential falls into this class of
examples.

\subsubsection{4D Example 2, Time dependent or time independent Vector
potentials only}

In this example,the 1-form is presumed to be of the form 
\begin{equation}
A_{0}=A_{x}(x,y,z,t)dx+A_{y}(x,y,z,t)dy+A_{z}(x,y,z,t)dz,
\end{equation}
The time dependent potentials admit both a magnetic field $\mathbf{B}$ and
an electric field $\mathbf{E,}$ and the time independent potentials only
admit a $\mathbf{B}$ field.. \ The invariant Adjoint Jacobian technique
indicates that a charge current density is not induced, $J_{s}=0,$ and the
interaction energy is identically zero, $A\symbol{94}J_{s}=0$. \ The time
dependent potentials are of Pfaff dimension 4, and can support non-zero
Helicity and non-zero Parity without inducing a charge current density. \
The time independent potentials are Pfaff dimension 3, hence the topological
parity, $F\symbol{94}F,$ is zero and the closed integrals of topological
torsion have rational ratios. \ The Gaussian curvature and the mean
curvature are not necessarily zero, although the Adjoint curvature is always
zero; $A\symbol{94}J_{s}=0$. \ Hence the 3D hypersurface has degenerated
into a 2 dimensional surface in 4 dimensions.

\begin{eqnarray}
Mean &\neq &0, \\
Gauss &\neq &0 \\
Kubic &=&0.
\end{eqnarray}

\subsubsection{4D Example 3, Vector and Scalar Potentials without explicit
time dependence.}

In this example,the 1-form is presumed to be of the form 
\begin{equation}
A_{0}=A_{x}(x,y,z)dx+A_{y}(x,y,z)dy+A_{z}(x,y,z)dz-\phi (x,y,z)dt.
\end{equation}

The results of the formalism indicate that there is no spatial current
density, but there can exist a charge density. \ The adjoint curvature is
not necessarily zero, $A\symbol{94}J_{s}\neq 0$ .

\begin{eqnarray}
Mean &\neq &0, \\
Gauss &\neq &0 \\
Kubic &\neq &0.
\end{eqnarray}

The Pfaff dimension can be as high as 4, with the 1-form supporting both a
Helicity 3-form and a Parity 4 form.

\subsubsection{\protect\vspace{1pt}4D Example 4, Bateman - Whittaker
solutions}

In this example,the 1-form is presumed to be of the form 
\begin{equation}
A_{0}=\alpha (x,y,z,t)d\beta (x,y,z,t).
\end{equation}
The results of the formalism indicate that there can be both a current
density and a charge density, $J_{s}\neq 0$. \ The Pfaff dimension is 2 or
less. \ The Helicity 3 form and the Parity 4-form vanish. \ The $\mathbf{E}$
field and $\mathbf{B}$ field are always orthogonal. \ However, if the
function $\beta $ is independent from time (but $\alpha $ remains explicitly
time dependent) then the Adjoint curvature and the induced charge current
density also vanish. \ The 3D hypersurface reduces to a 2D hypersurface that
supports mean and Gauss curvature, but not cubic curvature invariants, $A%
\symbol{94}J_{s}=0$. 
\begin{eqnarray}
Mean &\neq &0, \\
Gauss &\neq &0 \\
A\symbol{94}J_{s}\;\;Kubic &=&0 \\
Top\_Torsion &\neq &0 \\
J_{s} &\neq &0
\end{eqnarray}
\ This result corresponds to what are called time harmonic solutions in the
engineering literature, and gives yet more credence to the idea that charge
current densities can be related to cubic curvatures.

\subsubsection{\protect\vspace{1pt}4D Example 5, A Hopf map solution. \ }

\vspace{1pt}In this example,the Hopf 1-form is presumed to be of the form 
\begin{equation}
A_{0}=b(ydx.-xdy)+a(tdz-zdt).
\end{equation}
The 1-form of Potentials depends on the coefficients $a$ and $b$ which are
presumed to take on values $\pm 1.$ \ \ There are two cases corresponding to
left and right handed ''polarizations'': \ $a=b$ or $a=\ -b$. \ What is
remarkable for this solution, is that both the mean curvature and the
Adjoint curvature of the implicit hypersurface in 4D vanish, for any choice
of a or b. \ The Gauss curvature is non-zero, positive real and is equal to
the square of the radius of a 4D euclidean sphere. \ The cubic interaction
energy density is zero. 
\begin{eqnarray}
Mean &=&0, \\
Gauss &>&0 \\
A\symbol{94}J_{s}\;\;Kubic &=&0 \\
Top\_Torsion &\neq &0 \\
J_{s} &\neq &0
\end{eqnarray}

\ This situation occurs when the three curvatures of the implicit 3-surface
are $\{0,0,+i\omega ,-i\omega ).$. \ The Hopf surface is a 3D imaginary 
\textit{minimal} surface in 4D and has two non-zero imaginary curvatures! \
Strangely enough the charge-current density is not zero, but it is
proportional to the topological Torsion vector that generates the 3 form $A%
\symbol{94}F.$ \ The topological Parity 4 form is not zero, and depends on
the sign of the coefficients a and b. \ In other words the 'handedness' of
the different 1-forms determines the orientation of the normal field with
respect to the implicit surface. \ This set of circumstances corresponds to
the Case 3 \ situation described above where the charge current interaction
density is zero, but the charge current density is not zero. \ It is known
that a process described by a vector proportional to the topological torsion
vector in a domain where the topological parity is non-zero is
thermodynamically irreversible [10]. \ 

This example demonstrates that in special cases the charge-current density
is not proportional to the adjoint curvature of the implicit minimal
surface. \ However, the case corresponds to the special situation where the
interaction energy, alias the internal energy relative to the given process,
is zero, and yet the process is non-zero, but thermodynamically irreversible.

\section{Summary and applications to p-brane theories}

Every Pfaffian 1-form whose coefficients are functionally homogeneous of
degree zero can be used to describe the normal field to an implicit surface.
\ The curvature similarity invariants can be computed from the Jacobian
matrix of the homogeneous 1 form. \ \ For those p-branes which are 3
dimensional implicit surfaces in 4 dimensions, it is possible to deduce an
electromagnetic interpretation with an intrinsic charge current density. \
The interaction energy density of this charge current density and the
potentials that define the implicit surface is exactly the cubic curvature
similarity invariant of the implicit hypersurface. \ As the curvature radii
get smaller and smaller, the electromagnetic interaction energy being
proportional to the cube of the curvatures could conceivably prevent if not
impede gravitational collapse. \ Certainly such terms should be included in
the dynamics of collapsing mass systems. \ \ This effect, like the
Bohm-Aharanov effect, does not depend explicitly upon the field strengths, $%
\mathbf{E}$ and $\mathbf{B}$. \ \ Such a possibility appears to have been
neglected in metric based curvature theories.

\bigskip

\section{Acknowledgments}

This work is dedicated to Pucky (alias Elie Cartan), a constant and devoted
companion, who died on December 7, 2000.

\smallskip

\section{References}

\lbrack 1\rbrack\ The importance of the N-1 form A\symbol{94}H (now written
as A\symbol{94}G) was first anticipated in: R. M. Kiehn and J. F. Pierce, 
\textit{An Intrinsic Transport Theorem} Phys. Fluids 12, 1971. \ The concept
was further employed in R. M. Kiehn, \textit{Periods on manifolds,
quantization and gauge}, J. of Math Phys 18 no. 4 p. 614 (1977), and with
applications described in R. M. Kiehn, \textit{Are there three kinds of
superconductivity,} Int. J. Mod. Phys B 5 1779. (1991)

\lbrack 2\rbrack\ R. M. Kiehn, \textit{The Photon Spin and other Topological
Features of Classical Electromagnetism} in Vigier2000, edited by V. Amoroso
and J. Hunter, \ ((Kluwer, Dordrecht 2001 to appear)

R. M. Kiehn, \textit{Topological Evolution of classical electromagnetic
fields and the photon}, in ''The Photon and the Poincare Group\textit{''},
edited by V. Dvoeglazov, (Nova Science Publishers, NY 1999)

\lbrack 3\rbrack\ E. J. Post, ''Quantum Reprogramming\textit{''}, \ (Kluwer,
Dordrecht 1995)

[4]\ \ R. M. Kiehn, \textit{Topological Torsion and Topological Spin as
Coherent Structures in Plasmas}

\ \ \ \ \ http://www22.pair.com/csdc/pdf/plasma3.pdf

\lbrack 5\rbrack\ M. Gockler and T, Schucker, ''Differential geometry, gauge
theories and gravity'', Cambridge University Press, (1990) p.44-45.

\lbrack 6\rbrack\ H. Bateman, \textit{Electrical and Optical Wave Motion},
(Dover, New York, 1914, 1955) p.12.

\lbrack 7\rbrack\ L. Brand, ''Vector and Tensor Analysis'', (Wiley NY, 1962)
p. 150.

\lbrack 8\rbrack\ Chern, S. S., ''A Simple intrinsic proof of the Gauss
-Bonnet formula for Closed Riemannian Manifolds'', Annals of Math. 45,
(1944), p. 747- 752. \ also see Chern, S.S. , ''Historical Remarks on
Gauss-Bonnet'', MSRI 04808-88, Berkeley, CA (1988).

[9]\ R. M. Kiehn, \textit{Holder Norms (3D),}

http://www22.pair.com/csdc/pdf/holder3d.pdf

\ \ \ \ \ \ R. M. Kiehn, \textit{Holder Norms (4D),}

\ http://www22.pair.com/csdc/pdf/holder4d.pdf

\ \ \ \ \ \ R. M. Kiehn, \textit{Implicit surfaces,}

http://www22.pair.com/csdc/pdf/implinor.pdf

[10]\ R. M. Kiehn, \textit{Dissipation, Irreversibility and Symplectic
Lagrangian} \textit{Systems on Thermodynamic Space of Dimension 2n+2,}

\textit{\ \ }http://www22.pair.com/csdc/pdf/irrev1.pdf

\vspace{1pt}

\begin{quote}
\vspace{1pt}
\end{quote}

\end{document}